\documentclass[conference]{IEEEtran}
\IEEEoverridecommandlockouts

\usepackage{cite}
\usepackage{amsmath,amssymb,amsfonts}
\usepackage{algorithmic}
\usepackage{graphicx}
\usepackage{textcomp}
\usepackage{xcolor}
\usepackage{soul}
\usepackage{booktabs}
\usepackage{url}
\usepackage{tabularx}
\usepackage{multirow}

\def\BibTeX{{\rm B\kern-.05em{\sc i\kern-.025em b}\kern-.08em
    T\kern-.1667em\lower.7ex\hbox{E}\kern-.125emX}}
\begin{document}

\title{Users and Wizards in Conversations:\\How WoZ Interface Choices Define Human-Robot Interactions}

\author{\IEEEauthorblockN{Ekaterina Torubarova}
\IEEEauthorblockA{\textit{Division of Speech, Music and Hearing} \\
\textit{KTH Royal Institute of Technology}\\
Stockholm, Sweden \\
ekator@kth.se}
\and
\IEEEauthorblockN{Jura Miniota}
\IEEEauthorblockA{\textit{Division of Speech, Music and Hearing} \\
\textit{KTH Royal Institute of Technology}\\
Stockholm, Sweden \\
jura@kth.se}
\and
\IEEEauthorblockN{Andre Pereira}
\IEEEauthorblockA{\textit{Division of Speech, Music and Hearing} \\
\textit{KTH Royal Institute of Technology}\\
Stockholm, Sweden \\
atap@kth.se}
}

\maketitle

\begin{abstract}
In this paper, we investigated how the choice of a Wizard-of-Oz (WoZ) interface affects communication with a robot from both the user's and the wizard's perspective. In a conversational setting, we used three WoZ interfaces with varying levels of dialogue input and output restrictions: a) a restricted perception GUI that showed fixed-view video and ASR transcripts and let the wizard trigger pre-scripted utterances and gestures; b) an unrestricted perception GUI that added real-time audio from the participant and the robot c) a VR telepresence interface that streamed immersive stereo video and audio to the wizard and forwarded the wizard's spontaneous speech, gaze and facial expressions to the robot. 
We found that the interaction mediated by the VR interface was preferred by users in terms of robot features and perceived social presence. For the wizards, the VR condition turned out to be the most demanding but elicited a higher social connection with the users. VR interface also induced the most connected interaction in terms of inter-speaker gaps and overlaps, while Restricted GUI induced the least connected flow and the largest silences. Given these results, we argue for more WoZ studies using telepresence interfaces. These studies better reflect the robots of tomorrow and offer a promising path to automation based on naturalistic contextualized verbal and non-verbal behavioral data.
\end{abstract}


\section{Introduction}
In the field of Human-Robot Interaction (HRI), Wizard-of-Oz (WoZ) interfaces where a human operator replaces part or the entire decision process of a system, are commonplace \cite{riek2012wizard}, especially in scenarios involving complex social decision-making or language exchange. Many HRI studies employ WoZ settings, with humans controlling one or more robot functions \cite{baxter2016characterising}. Yet, the wizard's data from these studies, rich in potential, is often discarded post-study, however, this data can be invaluable for different applications. Immersive WoZ interfaces offer a unique opportunity in this regard as they allow for the collection of rich, context-specific data that can be used in refining behaviors and decision-making processes.


\begin{figure}[t]
    \centering
    \includegraphics[width=0.9\linewidth]{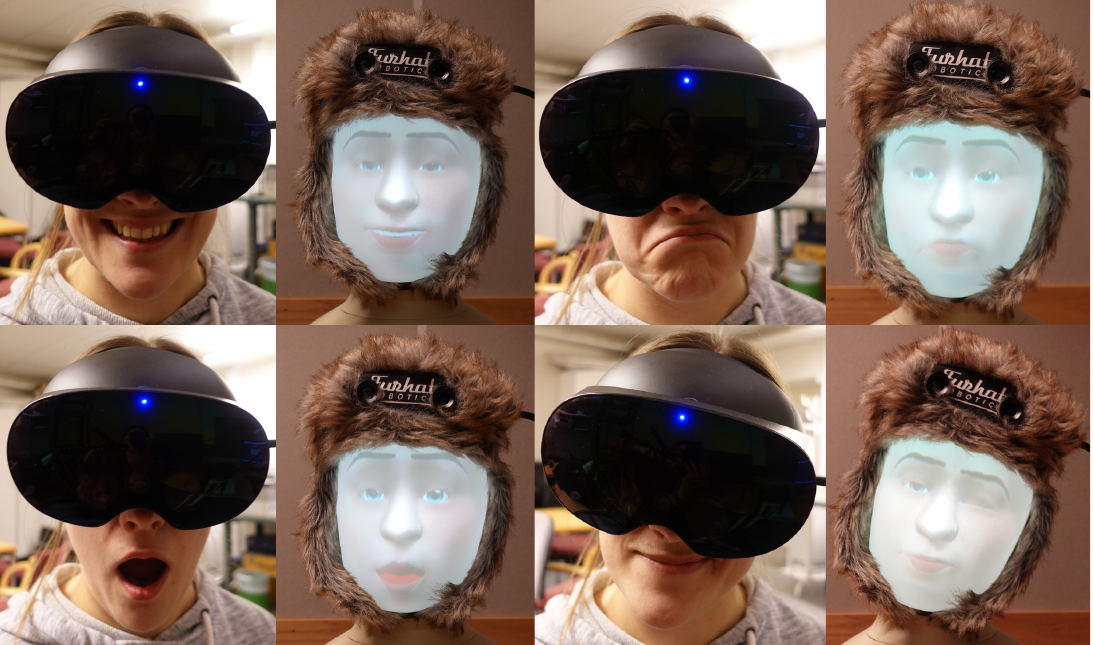}
    \caption{An example of the Meta Quest Pro tracking the operator's facial expression and displaying them on the Furhat robot in real-time.}
    \label{fig:face_tracking}
\end{figure}

In this paper, we delve into a comparative analysis of three different WoZ interfaces, each offering varying degrees of input and output capabilities that will result in varying levels of flexibility and control: a) a restricted perception graphical user interface (GUI) receiving a fixed-view video stream and automatic speech recognition (ASR) feedback from the participants, with pre-written utterances and gestures delivered via synthetic robot voice; b) an unrestricted perception GUI receiving fixed-view video, and real-time audio from the participant, offering the same output; and c) a VR interface receiving immersive video and real-time audio, with unrestricted spontaneous natural speech and gestures. Our experiment involved 25 users and 5 novice wizards engaging in 5-minute human-robot conversations centered around ethical dilemmas. Each wizard, controlling the robot via each of the three interfaces, engaged in three conversations with 5 unique users in a within-subjects design, resulting in 15 interactions per wizard (75 in total). 

We hypothesize that these interface variations will influence both the communication efficiency and users' preferences for the robot. In particular, we expect that the interaction mediated by the immersive VR interface will foster deeper engagement and perceived social connection, enabling more natural conversations due to enhanced turn-taking and non-verbal expression capabilities. Additionally, by assessing the workload and habituation process across the three interfaces of the wizards who do not have prior experience operating robots, we hypothesize the VR interface to provide the most intuitive control resembling real-life human-human interaction. 

The unique contributions of this paper include 1) testing and sharing a novel VR WoZ interface compared to more common GUI interfaces; 2) analysis of both users' and wizards' data, lacking in prior studies on WoZ methodologies; 3) in-depth qualitative assessment of novice wizards' experience of acquiring different interfaces. Our findings aim to guide future decisions on WoZ interface design, whether for targeted HRI studies or paving the path for automation.

\section{Related Work} 
\subsection{WoZ in Conversational Systems and HRI}
Multiple studies investigated the intricacies of the WoZ approach in tasks requiring verbal communication with the user due to its complexity and sensitivity \cite{riek2012wizard}. A foundational paper delved into the potential of the WoZ technique to emulate interactive technologies, with a special emphasis on speech input and output systems \cite{fraser1991simulating}. More than 30 years later, implementing believable verbal communication still remains a challenge: for instance, with the prevalence of Large Language Models, recent research suggests deploying them as language 'Scarecrows' - quick and capable, but requiring cautiousness and transparency \cite{williams2024scarecrows}.   

In verbal communication, there is a great demand for the wizard to rapidly select and input open-domain content. The study by \cite{hu2023wizundry} tackled the problem of a single wizard load by proposing a cooperative multi-wizard platform Wizundry, where the wizards could split speech-based tasks in real-time, coming up with a collaborative strategy. The study by \cite{Aylett_2020} proposed to address this challenge through 'voice puppetry'. Their system leverages a neural TTS system in near real-time, allowing natural speech input to determine the output for a specific synthetic target voice. 
With a high demand on wizard, in an extended conversation, turn-taking timing when producing a behavior is particularly important: response latency in everyday communication is reported to be universally as short as 200 ms \cite{stivers2009universals}. While a significant line of research is dedicated to modeling turn-taking in conversational systems, fewer works explicitly analyze WoZ turn-taking. A study by \cite{devault2015toward} compared their prototypical WoZ dialogue corpus to face-to-face (FTF) dialogue. They found longer speaker switch time in WoZ dialogues and more overlapped speech in FTF dialogues, suggesting tighter turn coordination, but no difference in speech duration between speakers. Delayed turn-taking has also been consistently linked to lower engagement \cite{pellet2023multimodal}. Thus, we consider it valuable to investigate both the turn-taking dynamics of WoZ dialogues and the subsequent impressions of end users and wizards. 

Another early study argued that computer dialogue interfaces shouldn't merely replicate human-human interactions. Instead, they should be anchored in the distinct characteristics inherent to human-computer interactions \cite{dahlback1993wizard}. Building on this foundation, a more recent exploration into the dynamics between humans and various communicative agents, such as chatbots, voice assistants, and social robots, pinpointed the unique challenges each agent presents \cite{greussing2022researching}. This underscores the significance of robust WoZ simulations specialized for HRI. A preliminary study with a humanoid robot \cite{shin2019apprentice} observed that wizards experience difficulties in four different sub-tasks: attention, decision, execution, and reflection. A detailed exploration into WoZ dialogues \cite{funakoshi2019toward} further scrutinized the challenges of emulating natural human dialogues using a Graphical User Interface (GUI) WoZ approach. While the interactions were enjoyable for the users, this study highlights the difficulties of fully incorporating an operator’s intricate interaction tactics into dialogue systems with limited dialogue options. A recent study \cite{song2022costume} examined how the balance between robotic features and wizard's features affects the perception of teleoperated robots in a supermarket setting. This study examined degrees of social presence by varying the robot's interaction modes: with or without a wizard's face and with or without conversion to a robotic voice. The in-between condition, with the wizard's face and converted robotic voice, was perceived as more trustworthy and was able to maintain longer interactions with customers. This study highlights the delicate balance between anthropomorphizing robots and maintaining their inherent robotic nature for optimal user engagement in social interaction scenarios.

\subsection{Virtual Reality (VR) WoZ for Immersive Robot Telepresence}
In the domain of Immersive Robot Telepresence (IRT), the design and efficiency of remote control interfaces play a crucial role in determining operator performance for remotely controlled robots. Multiple studies showed that VR-mediated interfaces can allow high immersivity and efficient, intuitive task execution. An insightful study compared three remote control interfaces for mobile inspection robots, particularly emphasizing the potential of VR devices \cite{jankowski2015usability}. The primary control interface incorporated a VR headset, data gloves for enhanced gripper control, joystick-based movement controls, and a motion tracking system to gauge head orientation and hand positions. The results showed that VR-enhanced interfaces significantly boosted operator productivity, heightened spatial presence, and improved distance evaluation. Notably, such VR interfaces also reduced the adaptation time for operators, attributing this efficiency to heightened intuitive control and comfort levels. In \cite{thellman2017like}, an operator wore a VR headset and used handheld controllers to control a Pepper robot by simply moving their body, allowing for intuitive embodied acting and efficient robot control with first-person view. The authors claimed their system could reduce wizard response times and produce more contextually apt robot behaviors. However, the system did face challenges such as motion sickness and lagging movements caused by the slow servos present in the Pepper robot. They also underscored the need for future integration of eye-tracking technology, which would elevate the realism and responsiveness of the robot interaction in social contexts. Similarly, \cite{tran2018hands} utilized a VR headset connected to the robot's camera perspective, coupled with a Leap Motion Controller that offered the operator a hands-free interaction medium to control the Pepper robot. An augmented reality interface was introduced to facilitate the control of a robot's verbal and nonverbal behaviors in \cite{pereira2017augmented}. The results showed that when a single operator controlled both types of behavior that require tight coordination, it yielded better task results and provided a heightened sense of 'presence' compared to when the task was split between two operators (one for verbal and another for nonverbal behavior). The study also suggested that this interface, which harnesses full-body motion control for the robot while also enabling verbal interactions, presented a powerful tool for WoZ interactions in human-robot scenarios, as well as for data collection to inform the development of autonomous multimodal dialogue systems.

\subsection{WoZ for Data Collection or Automation}
WoZ interfaces provide valuable data for learning models of social behavior and have been used to automate robot actions. In \cite{knox2016learning}, the Learning from the Wizard approach was proposed, where robots learn a behavior policy from WoZ demonstrations. This study showed that robots that learned from prior interactions successfully engaged children in play. In \cite{sequeira2016discovering}, the authors used WoZ interactions with constrained wizard perception, yielding a socially aware and empathetic collaborative robot tutor. Supervised Progressively Autonomous Robot Competencies was proposed in \cite{senft2015sparc, senft2019teaching}, enabling robots to learn online from wizard guidance and become more autonomous. Another study \cite{lala2018evaluation} used an LSTM model on WoZ data to identify conversation shift moments, finding that model performance varied across contexts, highlighting the importance of tailoring WoZ data collection. The authors of \cite{engwall2022wizard} explored human-to-robot conversation transition for language learning, showing high accuracy in wizard response selection, transferable to the robot, hinting at a path toward automation. While our paper doesn't focus on automation, the proposed telepresence system can facilitate the collection of more naturalistic data for future automation endeavors.

\section{System}

\begin{table}[]
\resizebox{\linewidth}{!}{%
\begin{tabularx}{\linewidth}{@{}Xll@{}}
\toprule
\textbf{Interface} & \textbf{Input Type} & \textbf{Output Type} \\ \midrule
\begin{tabular}[c]{@{}l@{}}Restricted Perception GUI\end{tabular} & 
\begin{tabular}[c]{@{}l@{}}
\textbullet\ fixed-view video\\
\textbullet\ ASR transcriptions
\end{tabular} & 
\begin{tabular}[c]{@{}l@{}}
\textbullet\ pre-written speech\\(synthesized voice)\\
\textbullet\ built-in gestures
\end{tabular} \\
\hline
\begin{tabular}[c]{@{}l@{}}Unrestricted Perception GUI\end{tabular} & 
\begin{tabular}[c]{@{}l@{}}
\textbullet\ fixed-view video\\
\textbullet\ real-time audio
\end{tabular} & 
\begin{tabular}[c]{@{}l@{}}
\textbullet\ pre-written speech\\(synthesized voice)\\
\textbullet\ built-in gestures
\end{tabular} \\
\hline
VR Telepresence Interface& 
\begin{tabular}[c]{@{}l@{}}
\textbullet\ immersive video\\
\textbullet\ real-time audio
\end{tabular} & 
\begin{tabular}[c]{@{}l@{}}
\textbullet\ unrestricted speech\\(natural voice)\\
\textbullet\ spontaneous gestures
\end{tabular} \\ \bottomrule \addlinespace[2pt]
\end{tabularx}%
}
\caption{Comparison of the input and output of each interface.}
\label{tab:conditions}
\end{table}

\begin{figure}[t]
    \centering
    \includegraphics[scale=0.35]{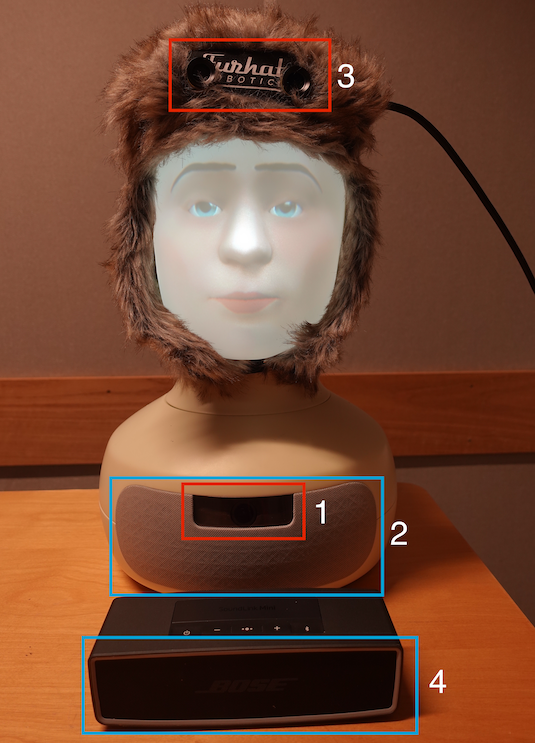}
    \caption{Furhat robot: 1) built-in static camera; 2) built-in speaker; 3) stereo camera; 4) external speaker.}
    \label{fig:furhat}
\end{figure}

\subsection{Social Robot}
In this study, we used the humanoid robotic head Furhat 2.7.0 \cite{al2012furhat}, capable of advanced speech synthesis and speech recognition (Fig. \ref{fig:furhat}). The robot's face is back-projected on the plastic mask, allowing for different appearances, believable eye-gaze and realistic facial movements (for lip-synchronization and timely expressions that reveal emotional states such as sadness and excitement). The motors in the robot's neck allow for head movements such as nods or headshakes. Additionally, it is equipped with a large voice library, comprising both male and female voices.

\subsection{Graphical User Interfaces}
\begin{figure}[t]
    \centering
    \includegraphics[width=\linewidth]{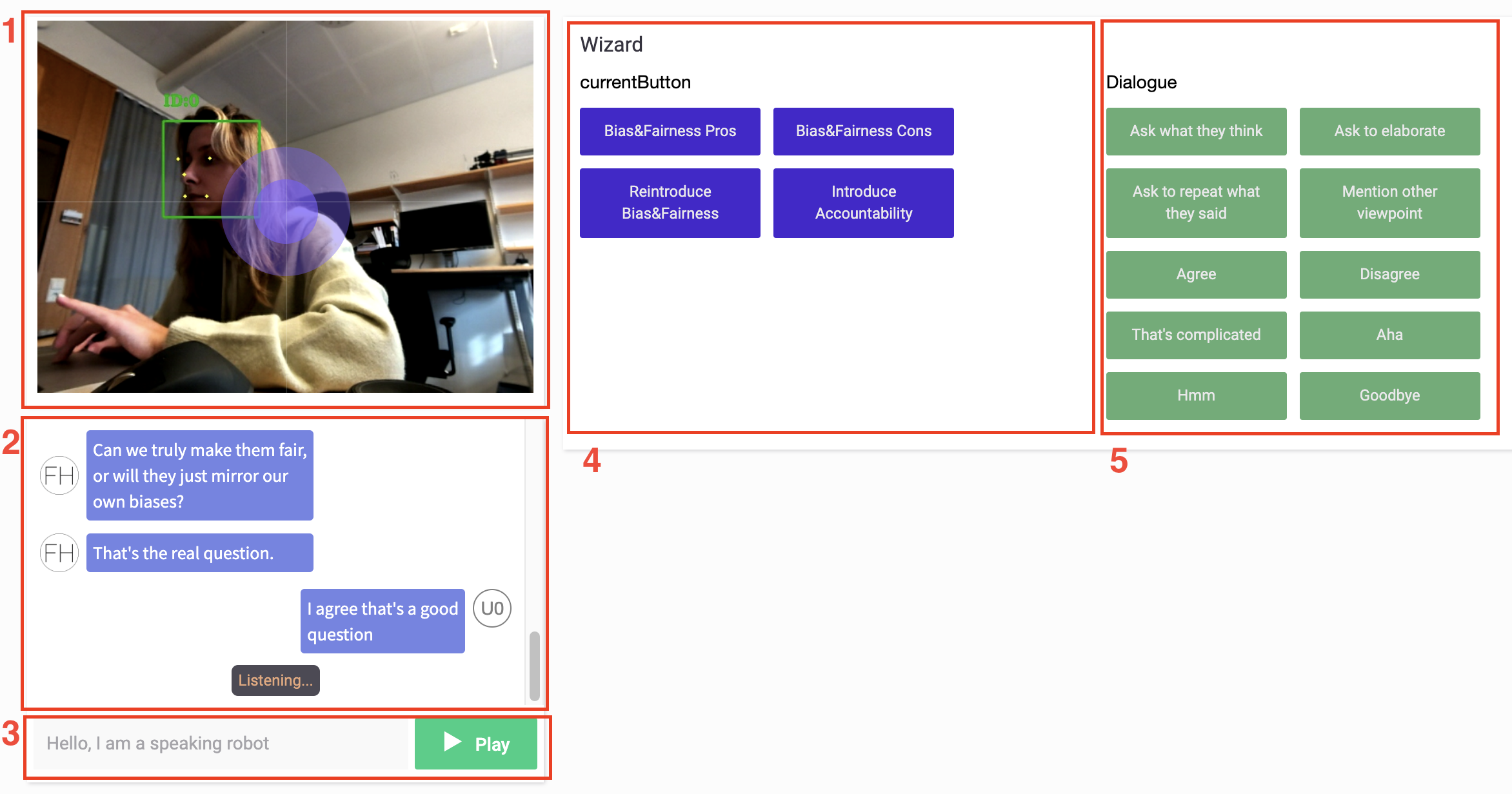}
    \caption{WoZ GUI in the middle of the conversation: 1) robot's camera view; 2) dialogue history; 3) typing box; 4) dynamic buttons; 5) static buttons.}
    \label{fig:dashboard}
\end{figure}

For the GUI-based interfaces, we used the built-in web interface for wizarding the Furhat robot (see Fig. \ref{fig:dashboard}). By default, this interface offers the wizard a video stream from the fixed-view robot's camera (Fig. \ref{fig:dashboard}.1) and a dialogue history component that displays the utterances of both interlocutors, with the user speech recognized by the built-in ASR (Fig. \ref{fig:dashboard}.2). We employed this default interface as the \textit{Restricted GUI} condition. For the \textit{Unrestricted GUI} condition, we used the same system but added a real-time audio stream from the participant, sourced from a lapel microphone. 

We also created two sections of buttons corresponding to the robot's utterances: dynamic (Fig. \ref{fig:dashboard}.4) and static (Fig. \ref{fig:dashboard}.5). The dynamic section, in which the options changed based on the conversation state, had a maximum of four buttons displayed at any given time. In contrast, the static buttons remained consistently visible on the screen, with ten buttons in total: 1) ask the participant's opinion, 2) ask them to elaborate their statement, 3) ask the participant to repeat, 4) mention another viewpoint, 5) agree, 6) disagree, 7) give an uncertain response, 8-9) provide backchannels, 10) end the conversation. Each static button had multiple utterance options to ensure variety. This design ensured that pressing the same button repeatedly would not produce repetitive utterances. However, we maintained a sequential order for the different utterances linked to each static button to ensure a more consistent experience across users. In addition to the dialogue buttons, by default, the Furhat interface includes a typing box for producing a custom utterance (Fig. \ref{fig:dashboard}.3). The wizards were informed that they could use this option during the interaction but they should avoid long delays caused by typing.

\subsection{VR Telepresence Wizard Interface}
To implement the \textit{VR Interface} condition, we created a telepresence platform for the Furhat robot using the Meta Quest Pro\footnote{\url{https://www.meta.com/se/en/quest/quest-pro/}}, a VR headset that has built-in eye and face tracking. The facial expressions, head, and eye movements of the wizard wearing the headset were collected from the headset and displayed on the Furhat robot's mask in real-time (Fig. \ref{fig:face_tracking}). We used Meta platforms Movement SDK for Unity\footnote{\url{https://developer.oculus.com/documentation/unity/move-overview/}} and a Furhat CSharp interface inside the Unity\footnote{\url{https://unity.com/}} editor to integrate the VR headset with the Furhat robot to create a VR-mediated WoZ interface. The full open-source version of our code can be found in Github repository\footnote{\url{https://github.com/andre-pereira/Furhat-Telepresence-Zed-MetaQuestPro}}. 

While Furhat is equipped with a built-in camera at its base, it was not suitable for the VR setup. It was the fixed-view camera used in the initial two conditions, thus not allowing for a full immersion for the wizard. Instead, we used a stereo camera inserted into the robot's fur hat (Fig. \ref{fig:furhat}.3). This camera was located slightly above the eye level of the robot but oriented in a way that the wizard could keep the head straight and maintain eye contact with the user. It also allowed the wizard to move their head freely, maintaining a first-person view. The camera was directly connected to the computer to reduce lag and displayed the video to the operator in real time. Finally, wizard's speech was relayed to an external speaker (Fig. \ref{fig:furhat}.4) instead of the built-in speaker (Fig. \ref{fig:furhat}.2) which can only be used for Furhat's speech synthesis.

\subsection{Ethical Dilemmas Conversation Scenario}
As a conversational scenario, we chose three ethical dilemmas that we expected to provide an engaging 5-minute conversation. The dilemmas chosen for this study were
shared by the Stockholm Museum of Technology holding an exhibition on
AI and society prior to running this study. These dilemmas
were chosen for their relevance and ease of scripting, enabling participants to quickly memorize and navigate dialogues. The dilemmas were the following: 1) Should we have robot judges that use machine learning to make decisions? 2) If we invent a pill that stops aging but only 5\% of people can use it, should it be allowed? 3) Would you take a DNA test before a date if the other person asks you? 

\begin{figure}[t]
    \centering
    \includegraphics[width=0.8\linewidth]{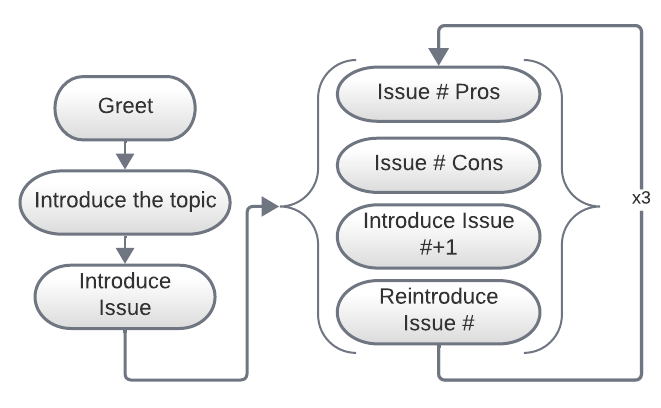}
    \caption{Dialogue script flow.}
    \label{fig:dialogue-flow}
\end{figure}

We developed a dialogue script for each of the topics which was used to generate the dynamic buttons in the GUIs (Fig. \ref{fig:dialogue-flow}). In the \textit{VR Interface} condition, the wizards were instructed to follow a similar structure of the dialogue but were not restricted in what they could say. The scenario for each topic included three issues (e.g. for the dilemma about robot judges: bias and fairness; accountability; and emotions). Each issue contained three pros and three cons. Apart from that, the wizards had the option to introduce the next issue or to reintroduce the current one again. To script the text of the utterances, we prompted ChatGPT\footnote{\url{https://chat.openai.com/}} with the following steps: 

\begin{enumerate}
    \item \textit{Imagine an ethical dilemma: [Dilemma Text]. Produce five main issues of [Question of the Dilemma].}
    \item \textit{Now provide three pros and three cons of [Issue] in this context.}
    \item \textit{Now write them in a more conversational way in one short paragraph.}
    \item \textit{Reintroduce [Issue] in three different ways as if the other person forgot what we were talking about, and we want to get back on track.}
\end{enumerate}

Each paragraph was further tuned to follow the topic more closely or provide more concrete examples and detailed descriptions. While refining the script, we selected three out of five issues raised about each topic to keep the conversation within the intended 5 minutes. To create a fairer comparison between our conditions we also added facial expressions available in Furhat's gesture library that were synchronized with the robot's speech whenever a button was pressed in both of our GUIs. For this, we prompted Gemini\footnote{\url{https://gemini.google.com/}} with the following: 

\textit{There is a script for what a social robot should say during a dialogue: [Script]. There is a list of facial expressions and gestures the robot can perform: BigSmile, Blink, BrowFrown, BrowRaise, CloseEyes, ExpressAnger, ExpressDisgust, ExpressFear, ExpressSad, GazeAway, Nod, Oh, OpenEyes, Roll, Shake, Smile, Surprise, Thoughtful, Wink. Select and distribute the gestures across the script where appropriate.
}

The result was refined in cases if gestures were too frequent or too sparse, or if they significantly interrupted the speech flow. We selected the best-performing language model for each of these tasks. Although the results were largely satisfactory, both the text of the script and the gestures were manually revised.

\section{Method}
\subsection{Participants}
We aimed at a sample size of 5 wizards, each interacting with 5 unique users across three conditions, resulting in a total of 75 interactions. 
The data of one wizard and all corresponding users was excluded from the analysis due to technical issues with the VR headset that occurred with 4 users. To reach our intended sample size, we recruited another wizard and 5 new users. All were recruited through university channels and a local experiment recruitment portal.

In the final analysis, 5 wizards were included (1 male, 4 female, age M = 29.4, SD = 3.5). Wizards' average familiarity with robots was medium to low (M = 2.2 out of 5, SD = 1.09), as well as average familiarity with VR (M = 2.6 out of 5, SD = 1.14). 4 out of 5 wizards reported no prior experience with social robots (one wizard has seen or interacted with Furhat and Pepper). All wizards had experience in performing arts (dance, stand-up comedy, theater, public speaking) and were native or fluent in English. The performing experience criterion was added to make sure the wizards were comfortable in the task of performing the robot's role. Prior to the experiment, they were interviewed and informed about the tasks they would have to do during the experiment. The wizards were rewarded with a gift card equal to 91 USD for their participation. 

As for the users, we analyzed data from 25 participants (11 male, 13 female, and 1 other; age min = 19, max = 57, M = 28.5, SD = 11.27). They had a low average familiarity with robots (M = 1.92 out of 5, SD = 0.79), and 65.3\% reported no prior experience with social robots. All users were fluent or native in English. They were rewarded with either a gift card equal to 9 USD or a cinema ticket for their participation. All of the participants gave written consent to participate in the study. According to the Swedish national regulations, ethical approval for the study was not required because we did not collect sensitive personal data or conduct physical intervention.

\subsection{Measures}
To evaluate the users' perception of the robot, we used the Godspeed questionnaire \cite{godspeed}. To evaluate the social aspects of the interaction, we used the social presence questionnaire described in \cite{pereira2020effects} evaluated in other social HRI settings. The questionnaire consists of 18 questions assessing six dimensions: co-presence, attention allocation, message understanding, behavioral interdependence, affect understanding, and affect interdependence. To evaluate the wizards' proficiency and progression experience, we used the NASA-TLX questionnaire \cite{hart2006nasa}, a common tool for assessing task load; and the System Usability Scale (SUS) \cite{brooke1996sus}. In the SUS questionnaire, we omitted questions 1 and 4 because they were not relevant for the current experiment, since they refer to future recurrent use of the systems, which we did not intend. 

\subsection{Experimental Setup}
The experiment was conducted in two adjacent rooms: the control room and the experimental room. The experimental room was soundproof with no visual connection to the control room. The robot was installed in the experimental room, and users sat in front of the robot at a comfortable distance and at the same eye level. In the the \textit{Restricted GUI} and \textit{Unrestricted GUI} conditions, the robot used its built-in static camera (Fig. \ref{fig:furhat}.1) and provided tracking software to autonomously attend to the user with eye and head movements, and had micro-expressions enabled (e.g. blinking, small head movements). In the \textit{VR Interface} condition, both attention and facial expressions were captured from the wizard's natural behavior that was displayed on the robot. The robot's appearance and voice were changed between conditions. Six robot mask textures were selected, three male and three female, and two male and female Amazon Polly voices from the built-in voice library. The order of the topics, interfaces, and robot features (corresponding to the wizard's gender) was randomized for each user. 
Video and audio data from the interaction were recorded with two cameras: one facing the user and another facing the robot.

\subsection{Procedure}
Before starting the experiment, the wizards filled in a questionnaire that reported on their familiarity with social robots and VR. Then they had a 1-hour practice to get used to the interfaces and the topics. They were trained on the functionalities of each interface and practiced using them with the experimenter. They were informed that the user would not know about them controlling the robot and that their task was to make the robot have a naturalistic and smooth conversation with the user. They were also asked to get familiar with the script for each topic. 

The wizard then left the room while a user entered and re-entered when the user was sitting in the experimental room. This procedure ensured that the wizard and the user did not see each other before the experiment. Before starting the interaction, the users also answered how familiar they were with social robots. Then they could read the topics they were supposed to discuss and think about their opinion. Next, they were instructed to have a free conversation with the robot about the topic that the robot would suggest for 5 minutes. After 5 minutes, the robot informed the participants that it was the time to wrap up the discussion. After the conversation, the users filled in the Godspeed and Social Presence questionnaires regarding that interaction. After that, they were instructed to have another conversation with the robot. They were told that the robot would be powered by a different system which would not have any recollection of their previous conversation. The same procedure was repeated for each of the three topics. Before and after each interaction, the robot's face projector was turned off and the user was told that the robot system was being changed. At the end of their experiment, the users were asked to answer two open-ended questions: (1) Did you feel any differences between the systems? If so, what kind of differences? (2) Did you feel that any of the systems could be controlled by a human? If so, which aspects of which system?

For each interaction, the wizards were reminded about the topic and the interface they would use. They had time to check the script if they needed to before the interaction. They were instructed to have a conversation for 5 minutes and finalize it by saying goodbye to the user when appropriate once the experimenter signaled to them by a gentle tap on the back that the time had run out. After each interaction, the wizards filled in the NASA-TLX questionnaire. 

After discussing all three topics and filling out the questionnaires, the users were debriefed that the robot was in fact controlled by a human operator and they could meet the wizard. At the end of the experiment day, the wizards filled in a System Usability Scale (SUS) for each of the systems, and left their open-ended comments about their experience. After they were done filling in all questionnaires they were asked to sum up their experience in a semi-structured manner.
\section{Results}
\subsection{User Questionnaires}
 \subsubsection{Godspeed} To examine whether Godspeed scores were affected by experimental condition, we first calculated a mean score for each section of the questionnaire as well as the mean total score for each participant. Then, we fitted a linear mixed-effects model (LMM) with condition as a fixed effect and wizard ID and participant ID as random effects to account for potential variability between wizards and individual users. For each section as well as the total score, the \textit{VR Interface} condition achieved the largest mean score followed by \textit{Unrestricted GUI} followed by \textit{Restricted GUI}. Each comparison showed significant differences between conditions \((p < 0.05)\), except for the Safety score comparison between \textit{Restricted GUI} and \textit{Unrestricted GUI} \((p > 0.05)\). Full results for each comparison see in Table \ref{tab:godspeed}.
 


\begin{table*}[]
\resizebox{\textwidth}{!}{%
\begin{tabular}{lllllllll}
\toprule
\textbf{Godspeed Section} & \textbf{Condition} & \textbf{Estimate} & \textbf{SE} & \textbf{Comparison} & \textbf{Estimate} & \textbf{SE} & \textbf{t-value} & \textbf{p-value} \\ \midrule \addlinespace[1pt]
\multirow{3}{*}{Anthropomorphism} 
& Restricted GUI & 2.15 & 0.17 & Restricted GUI vs Unrestricted GUI & -0.6 & 0.18 & -3.35 & $\mathbf{<.01}$ \\
& Unrestricted GUI & 2.76 & 0.17 & Restricted GUI vs VR & -0.93 & 0.18 & -5.15 & $\mathbf{<.0001}$ \\
& VR & 3.69 & 0.17 & Unrestricted GUI vs VR & -1.54 & 0.18 & -8.51 & $\mathbf{<.0001}$ \\ \addlinespace[1pt] \cline{1-9} \addlinespace[1pt]
\multirow{3}{*}{Animacy} 
& Restricted GUI & 2.47 & 0.18 & Restricted GUI vs Unrestricted GUI & -0.63 & 0.18 & -3.43 & $\mathbf{<.01}$ \\
& Unrestricted GUI & 3.10 & 0.18 & Restricted GUI vs VR & -1.14 & 0.18 & -6.21 & $\mathbf{<.0001}$ \\
& VR & 3.61 & 0.18 & Unrestricted GUI vs VR & -0.51 & 0.18 & -2.78 & $\mathbf{<.01}$ \\ \addlinespace[1pt] \cline{1-9} \addlinespace[1pt]
\multirow{3}{*}{Likeability} 
& Restricted GUI & 2.88 & 0.16 & Restricted GUI vs Unrestricted GUI & -0.66 & 0.14 & -4.49 & $\mathbf{<.0001}$ \\
& Unrestricted GUI & 3.55 & 0.16 & Restricted GUI vs VR & -1.15 & 0.14 & -7.79 & $\mathbf{<.0001}$ \\
& VR & 4.04 & 0.16 & Unrestricted GUI vs VR & -0.48 & 0.14 & -3.30 & $\mathbf{<.01}$ \\ \addlinespace[1pt] \cline{1-9} \addlinespace[1pt]
\multirow{3}{*}{Intelligence} 
& Restricted GUI & 3.16 & 0.14 & Restricted GUI vs Unrestricted GUI & -0.45 & 0.13 & -3.35 & $\mathbf{<.01}$ \\
& Unrestricted GUI & 3.61 & 0.14 & Restricted GUI vs VR & -0.79 & 0.13 & -5.82 & $\mathbf{<.0001}$ \\
& VR & 3.95 & 0.14 & Unrestricted GUI vs VR & -0.33 & 0.13 & -2.47 & $\mathbf{<.05}$ \\ \addlinespace[1pt] \cline{1-9} \addlinespace[1pt]
\multirow{3}{*}{Safety} 
& Restricted GUI & 3.54 & 0.13 & Restricted GUI vs Unrestricted GUI & 0.02 & 0.13 & 0.20 & $n.s.$ \\
& Unrestricted GUI & 3.52 & 0.13 & Restricted GUI vs VR & -0.33 & 0.13 & -2.48 & $\mathbf{<.05}$ \\
& VR & 3.88 & 0.13 & Unrestricted GUI vs VR & -0.36 & 0.13 & -2.68 & $\mathbf{<.01}$ \\ \addlinespace[1pt] \cline{1-9} \addlinespace[1pt]
\multirow{3}{*}{\textbf{Total Score}} 
& Restricted GUI & 2.78 & 0.11 & Restricted GUI vs Unrestricted GUI & -0.50 & 0.11 & -4.51 & $\mathbf{<.0001}$ \\
& Unrestricted GUI & 3.29 & 0.11 & Restricted GUI vs VR & -1.05 & 0.11 & -9.30 & $\mathbf{<.0001}$ \\
& VR & 3.83 & 0.11 & Unrestricted GUI vs VR & -0.54 & 0.11 & -4.79 & $\mathbf{<.0001}$ \\ \addlinespace[1pt]
\bottomrule
\addlinespace[3pt]
\end{tabular}%
}
\caption{Linear mixed model results for each section of the Godspeed questionnaire across the three interface conditions. N.s. stands for p-value $>$ 0.05.}
\label{tab:godspeed}
\end{table*}

\subsubsection{Social Presence} For social presence, similarly to \cite{pereira2020effects}, we compared both the total mean score and a mean score for each section. Similarly to the analysis of Godspeed, we fitted an LMM with condition as a fixed effect and wizard ID and participant ID as random effects. For each section as well as the total score, the \textit{VR Interface} condition achieved the largest mean score followed by \textit{Unrestricted GUI} followed by \textit{Restricted GUI}. Each comparison showed significant differences between conditions \((p < 0.05)\), except for the Attention Allocation and Affect Interdependence score comparisons between \textit{Unrestricted GUI} and \textit{VR Interface} \((p > 0.05)\). Full results for each comparison see in Table \ref{tab:socialpresence}.


\begin{table*}[]
\resizebox{\textwidth}{!}{%
\begin{tabular}{lllllllll}
\toprule
\textbf{Social Presence Section} & \textbf{Condition} & \textbf{Estimate} & \textbf{SE} & \textbf{Comparison} & \textbf{Estimate} & \textbf{SE} & \textbf{t-value} & \textbf{p-value} \\ \midrule \addlinespace[1pt]
\multirow{3}{*}{Co-Presence} 
& Restricted GUI & 2.66 & 0.21 & Restricted GUI vs Unrestricted GUI & -0.70 & 0.25 & -2.80 & $\mathbf{<.001}$ \\
& Unrestricted GUI & 3.37 & 0.21 & Restricted GUI vs VR & -1.23 & 0.25 & -4.89 & $\mathbf{<.0001}$ \\
& VR & 3.90 & 0.21 & Unrestricted GUI vs VR & -0.52 & 0.25 & -2.09 & $\mathbf{<.05}$ \\ \addlinespace[1pt] \cline{1-9} \addlinespace[1pt]
\multirow{3}{*}{Attention Allocation} 
& Restricted GUI & 3.14 & 0.18 & Restricted GUI vs Unrestricted GUI & -0.73 & 0.24 & -2.97 & $\mathbf{<.01}$ \\
& Unrestricted GUI & 3.88 & 0.18 & Restricted GUI vs VR & -0.77 & 0.24 & -3.14 & $\mathbf{<.01}$ \\
& VR & 3.92 & 0.18 & Unrestricted GUI vs VR & -0.04 & 0.24 & -0.16 & $n.s.$ \\ \addlinespace[1pt] \cline{1-9} \addlinespace[1pt]
\multirow{3}{*}{Message Understanding} 
& Restricted GUI & 2.58 & 0.22 & Restricted GUI vs Unrestricted GUI & -0.69 & 0.27 & -2.56 & $\mathbf{<.05}$ \\
& Unrestricted GUI & 3.28 & 0.22 & Restricted GUI vs VR & -1.49 & 0.27 & -5.51 & $\mathbf{<.0001}$ \\
& VR & 4.08 & 0.22 & Unrestricted GUI vs VR & -0.80 & 0.27 & -2.85 & $\mathbf{<.01}$ \\ \addlinespace[1pt] \cline{1-9} \addlinespace[1pt]
\multirow{3}{*}{Behavioral Interdependence} 
& Restricted GUI & 2.41 & 0.17 & Restricted GUI vs Unrestricted GUI & -0.53 & 0.24 & -2.21 & $\mathbf{<.05}$ \\
& Unrestricted GUI & 2.94 & 0.17 & Restricted GUI vs VR & -1.04 & 0.24 & -4.19 & $\mathbf{<.0001}$ \\
& VR & 3.45 & 0.17 & Unrestricted GUI vs VR & -0.50 & 0.24 & -2.04 & $\mathbf{<.05}$ \\ \addlinespace[1pt] \cline{1-9} \addlinespace[1pt]
\multirow{3}{*}{Affect Understanding} 
& Restricted GUI & 1.96 & 0.17 & Restricted GUI vs Unrestricted GUI & -0.73 & 0.20 & -3.65 & $\mathbf{<.001}$ \\
& Unrestricted GUI & 2.69 & 0.17 & Restricted GUI vs VR & -1.18 & 0.20 & -5.91 & $\mathbf{<.0001}$ \\
& VR & 3.54 & 0.17 & Unrestricted GUI vs VR & -0.45 & 0.20 & -2.26 & $\mathbf{<.05}$ \\ \addlinespace[1pt] \cline{1-9} \addlinespace[1pt]
\multirow{3}{*}{Affect Interdependence} 
& Restricted GUI & 2.06 & 0.18 & Restricted GUI vs Unrestricted GUI & -0.41 & 0.20 & -2.02 & $\mathbf{<.05}$ \\
& Unrestricted GUI & 2.48 & 0.18 & Restricted GUI vs VR & -0.80 & 0.20 & -3.91 & $\mathbf{<.001}$ \\
& VR & 2.86 & 0.18 & Unrestricted GUI vs VR & -0.38 & 0.20 & -1.89 & $n.s.$ \\ \addlinespace[1pt] \cline{1-9} \addlinespace[1pt]
\multirow{3}{*}{\textbf{Total Score}} 
& Restricted GUI & 2.47 & 0.14 & Restricted GUI vs Unrestricted GUI & -0.63 & 0.18 & -3.52 & $\mathbf{<.001}$ \\
& Unrestricted GUI & 3.10 & 0.14 & Restricted GUI vs VR & -1.08 & 0.18 & -6.02 & $\mathbf{<.0001}$ \\
& VR & 3.56 & 0.14 & Unrestricted GUI vs VR & -0.45 & 0.18 & -2.50 & $\mathbf{<.05}$ \\ \addlinespace[1pt]
\bottomrule
\addlinespace[3pt]
\end{tabular}%
}
\caption{Linear mixed model results for each section of the Social Presence questionnaire across the three interface conditions. N.s. stands for p-value $>$ 0.05.}
\label{tab:socialpresence}
\end{table*}

\subsubsection{Perceived Autonomy}
In the open-ended responses regarding perceived autonomy, 38\% of users reported no differences between the conditions in the level of human control. 42\% suggested that the \textit{VR Interface} condition was controlled by a human. When specified why, 54\% of these users mentioned that the voice of the robot was coming from a human. This result contradicted both our pilot study with an expert wizard where no user reported awareness of human control \cite{Miniotaite_Torubarova_Pereira_2023}, and a study where participants' strong pre-existing perceptions about the robot's autonomy were resistant to change \cite{nasir2022questioning}. 

\subsubsection{Perceived Differences} 
In the open-ended responses by the users a common trend was either a gradient in preference between the conditions from \textit{Restricted GUI} as the least preferred to \textit{VR Interface} as the most preferred (e.g. \textit{"They were increasing in being humanlike and interactive."}; \textit{"I felt a very huge change in between the first interaction [Restricted GUI] and the last [VR Interface]. Felt it more organic in the third discussion."}); or a mention of the most preferred or the least preferred interaction (e.g.\textit{ "The last system [VR Interface] was more humanlike and enabled an organic conversation compared to the other two."}; \textit{"I preferred the last one [VR Interface]. I think this was due to the way the robot talked."}; \textit{"the last robot [Restricted GUI] was very machine-like. The second [VR Interface] was much human-like."}). 

The most common characteristics mentioned in the responses were fast vs. slow (12), human-like/organic vs. machine-like/artificial (7), high vs. low interactivity/responsiveness (5), little vs. a lot of interruptions (3), high vs. low attention (1), high vs. low emotion understanding (1), and best vs. worst voice (1).


\subsection{Wizard Questionnaires}
Given the limited sample size of the wizards, we focused on a qualitative data analysis rather than statistical inference.

\subsubsection{NASA-TLX}
To analyze the wizards' workload with different systems, we took an average of raw NASA-TLX scores for each wizard with each system per interaction. The average NASA-TLX score per system suggest that for our wizards, the VR interface was the most demanding to use (M = 4.04, SD = 1.22), followed by the Restricted GUI (M = 3.18, SD = 1.03), and Unrestricted GUI was the least demanding (M = 3.09, SD = 0.76). 


\subsubsection{System Usability Score}
To analyze the SUS for each system, we converted raw scores on a 1-5 Likert scale into a 0-100 scale \cite{brooke1996sus}. 
The results suggest that the Unrestricted GUI was the most usable (M = 62, SD = 11.2), followed by the VR interface (M = 43, SD = 19.6), and the Restricted GUI was the least usable (M = 38, SD = 14.3).


\subsubsection{Interviews}
In their open-ended reflections, W1 and W5 mentioned the Unrestricted GUI to be the easiest to use. 

W1, W5, and W6 noted the Restricted GUI to be frustrating to use due to difficulty in timing their responses and incorrect speech transcriptions. W2 and W4 mentioned that they felt rude when they interrupted the users with the Restricted GUI. W2 mentioned that using both GUIs \textit{"... was not so flexible to react to subtle changes in [the user's] flow"}. 

Regarding the VR interface, W2 and W5 mentioned that it was difficult for them to remember the scenario for the topics. W3 and W5 mentioned that they would prefer the Unrestricted GUI if they were not familiar with the topic and the VR interface if they were familiar with the topic. W2 wizard mentioned that they felt overwhelmed from concentrating using the VR interface because \textit{"I really tried to perform"}. W4 mentioned that it was inconvenient to use due to the echo they heard in their headphones. W1 mentioned that \textit{"the ability of the VR to translate the facial emotions made it the coolest to use". }

In terms of the social aspect, W5 mentioned that the VR interface was the most enjoyable: \textit{"... VR was more enjoyable for me personally. I had a larger degree of freedom and I was having a closer view of the participant which made the conversation run smoother after the second instance of VR use"}. W6 mentioned that \textit{"...with the VR, that I felt higher degree of responsibility"}. W4 and W5 highlighted the differences between the users they interacted with. W5: \textit{"If the participant showed reluctant and withdrawn, or defensive to the topics, I found the VR more challenging because it was in a greater degree up to me to build up rapport"}. W4 mentioned that with the Restricted GUI they could withdraw from the social awkwardness they felt with the audio and that it felt \textit{"easier socially"}.

In addition, W2 and W3 mentioned that they felt a difference between the condition most preferred by them and by users. W3: \textit{"I felt more comfortable when I couldn't hear them, but I imagine that they felt more comfortable during VR ... I didn't interrupt them ... But I had to socially engage more and it felt more demanding"}.

\begin{table*}[]
\resizebox{\textwidth}{!}{%
\begin{tabular}{lllllllll}
\hline
\textbf{Speech Parameter} & \textbf{Condition} & \textbf{Estimate} & \textbf{SE} & \textbf{Comparison} & \textbf{Estimate} & \textbf{SE} & \textbf{t-value} & \textbf{p-value} \\ \hline \addlinespace[1pt]
\multirow{3}{*}{\begin{tabular}[c]{@{}l@{}}Speech Duration (participant)\end{tabular}} 
& Restricted GUI & 92.90 & 8.36 & Restricted GUI vs Unrestricted GUI & -13.57 & 5.44 & -2.49 & $\mathbf{<.05}$ \\
& Unrestricted GUI & 106.47 & 8.36 & Restricted GUI vs VR & -30.55 & 5.27 & -5.52 & $\mathbf{<.001}$ \\
& VR & 123.45 & 8.42 & Unrestricted GUI vs VR & -16.98 & 5.52 & -3.07 & $\mathbf{<.005}$ \\ \addlinespace[1pt] \cline{1-9} \addlinespace[1pt]
\multirow{3}{*}{\begin{tabular}[c]{@{}l@{}}Speech Duration (robot)\end{tabular}} 
& Restricted GUI & 87.04 & 5.57 & Restricted GUI vs Unrestricted GUI& -23.56 & 4.79 & -4.91 & $\mathbf{<.001}$ \\
& Unrestricted GUI & 110.6 & 5.57 & Restricted GUI vs VR & 8.96 & 4.85 & 1.84 & $n.s.$ \\
& VR & 78.08 & 5.62 & Unrestricted GUI vs VR & 32.52 & 4.85 & 6.69 & $\mathbf{<.001}$ \\ \addlinespace[1pt] \cline{1-9} \addlinespace[1pt]
\multirow{3}{*}{\begin{tabular}[c]{@{}l@{}}Turn-Yielding Gaps Duration (between \\participant's and robot's utterance)\end{tabular}} 
& Restricted GUI & 6.70 & 0.35 & Restricted GUI vs Unrestricted GUI& 4.00 & 0.29 & 13.68 & $\mathbf{<.0001}$ \\
& Unrestricted GUI & 2.69 & 0.33 & Restricted GUI vs VR & 4.42 & 0.29 & 15.14 & $\mathbf{<.0001}$ \\
& VR & 2.27 & 0.33 & Unrestricted GUI vs VR & 0.42 & 0.26 & 1.57 & $n.s.$ \\ \addlinespace[1pt] \cline{1-9} \addlinespace[1pt]
\multirow{3}{*}{\begin{tabular}[c]{@{}l@{}}Turn-Processing Gaps Duration (between \\robot's and participant's utterance)\end{tabular}} 
& Restricted GUI & 1.98 & 0.11 & Restricted GUI vs Unrestricted GUI & 0.21 & 0.10 & 2.06 & $\mathbf{<.05}$ \\
& Unrestricted GUI & 1.76 & 0.10 & Restricted GUI vs VR & 0.29 & 0.10 & 2.71 & $\mathbf{<.001}$ \\
& VR & 1.69 & 0.10 & Unrestricted GUI vs VR & 0.07 & 0.09 & 0.75 & $n.s.$ \\ \addlinespace[1pt] \cline{1-9} \addlinespace[1pt]
\multirow{3}{*}{\begin{tabular}[c]{@{}l@{}}Cooperative Overlaps Proportion (robot's \\backchannels overlap with participant)\end{tabular}} 
& Restricted GUI & -4.41 & 0.56 & Restricted GUI vs Unrestricted GUI & -0.41 & 0.62 & -0.66 & $n.s.$ \\
& Unrestricted GUI & -3.37 & 0.44 & Restricted GUI vs VR & 0.01 & 0.68 & 0.01 & $n.s.$ \\
& VR & -4.18 & 0.54 & Unrestricted GUI vs VR & 0.42 & 0.58 & 0.74 & $n.s.$ \\ \addlinespace[1pt] \cline{1-9} \addlinespace[1pt]
\multirow{3}{*}{\begin{tabular}[c]{@{}l@{}}Destructive Overlaps Proportion \\(participant interrupts robot)\end{tabular}} 
& Restricted GUI & -4.17 & 0.61 & Restricted GUI vs Unrestricted GUI & 0.81 & 0.59 & 1.36 & $n.s.$ \\
& Unrestricted GUI & -4.98 & 0.68 & Restricted GUI vs VR & 0.90 & 0.64 & 1.39 & $n.s.$ \\
& VR & -5.07 & 0.71 & Unrestricted GUI vs VR & 0.08 & 0.69 & 0.12 & $n.s.$ \\ \addlinespace[1pt] \cline{1-9} \addlinespace[1pt]
\multirow{3}{*}{\begin{tabular}[c]{@{}l@{}}Destructive Overlaps Proportion \\(robot interrupts participant)\end{tabular}} 
& Restricted GUI & -1.53 & 0.24 & Restricted GUI vs Unrestricted GUI & 0.28 & 0.23 & 1.20 & $n.s.$ \\
& Unrestricted GUI & -1.82 & 0.23 & Restricted GUI vs VR & 1.24 & 0.30 & 4.16 & $\mathbf{<.0001}$ \\
& VR & -2.78 & 0.29 & Unrestricted GUI vs VR & 0.96 & 0.29 & 3.31 & $\mathbf{<.0001}$ \\ \addlinespace[1pt] \cline{1-9} \addlinespace[3pt]
\end{tabular}%
}
\caption{Linear mixed model results for comparing speech parameters across the three interfaces. Rows 1-4 show estimate in seconds, rows 5-7 show log-odds, n.s. stands for p-value $>$ 0.05.}
\label{tab:speech}
\end{table*}

\subsection{Speech Dynamics}
To analyze differences in speech dynamics across conditions, the audio recordings of the conversations were transcribed using WhisperX \textit{large-v2} model \cite{bain2022whisperx}. The transcriptions underwent a manual check, where speakers were also manually tagged. Since the model occasionally merged utterances divided by a long within-speaker pause into a single utterance, we used Silero Voice Activity Detector \cite{silero_vad_2024} with a 200 ms minimum silence threshold to obtain a more precise total speaker time. Since we were interested in conversational dynamics between the interlocutors, i.e. their involvement in the dialogue and efficiency of turn taking, we extracted the following speech parameters: speech duration of both speakers; duration of turn-yielding gaps (from participant to robot); duration of turn-processing gaps (from robot to participant); proportion of overlaps. 


\subsubsection{Speech Duration} To examine whether speech duration varied by condition and the speaker (participant or robot), we calculated the total speaker time for each interaction. We then fitted a linear mixed-effects model (LMM) separately for the participant's speech and for the robot's speech with condition as a fixed effect and wizard ID and participant ID as random effects to account for potential variability between wizards and individual users. The results of each following models see in Table \ref{tab:speech}. 

The participants spoke on average the longest in the \textit{VR Interface} condition, followed by the \textit{Unrestricted GUI}, and the least in the \textit{Restricted GUI} condition. All comparisons were significant \((p < 0.05)\). The robot, on the contrary, spoke on average the longest in the \textit{Unrestricted GUI} condition, and significantly reduced in both the \textit{Restricted GUI} and the \textit{VR Interface} conditions \((p < 0.001)\), although there was no significant difference between the latter two conditions \((p > 0.05)\).

\subsubsection{Turn-Yielding Gaps Duration}
For turn-yielding gap duration (the gaps between the participant's and the robot's utterances), we fitted an LMM to investigate the effect of the condition on the gap duration. 
The longest gaps were found in the \textit{Restricted GUI} condition, which significantly decreased in both \textit{Unrestricted GUI} and the \textit{VR Interface} condition \((p < 0.001)\). The comparison between the \textit{Unrestricted GUI} and the \textit{VR Interface} condition was not significant \((p > 0.05)\). 


\subsubsection{Turn-Processing Gaps Duration}
For turn-processing gaps duration (the gaps between the robot's and the participant's utterances), we fitted another LMM with the same parameters. 
The longest gaps were found in the \textit{Restricted GUI} condition, which significantly decreased in both \textit{Unrestricted GUI} and the \textit{VR Interface} condition \((p < 0.05)\). The comparison between the \textit{Unrestricted GUI} and the \textit{VR Interface} condition was not significant \((p > 0.05)\).


\subsubsection{Overlaps Proportion}
We aimed to differentiate between cooperative and disruptive overlaps. Several studies \cite{cafaro2016effects,hilton2016perception} found that cooperative overlaps, i.e. those indicating agreement, assistance, or clarification, are associated with higher engagement compared to disruptive overlaps, i.e. disagreement, floor taking, or topic change. Given the structure of our dialogue scenario, we focused on backchannels as instances of cooperative overlaps. Out of all overlaps, we manually tagged backchannels as cooperative overlaps and the other instances as disruptive. We investigated whether the condition affected the likelihood of overlaps occurring in an interaction, by calculating the proportion of each type of overlap from the total number of between-speaker transitions per interaction.

\paragraph{Cooperative Overlaps}
There were no instances of the participant's backchannels overlapping with the robot's utterance. For the opposite, i.e., the robot's backchannels overlapping with the participant's utterances, we fitted a binomial logistic regression model to investigate whether cooperative overlap proportion was affected by the condition. The model showed no significant effects of the condition \((p > 0.05)\). 

\paragraph{Destructive Overlaps}
We found no significant effect of the condition on the proportion of participants interrupting the robot \((p > 0.05)\). In the opposite case, we found a significant effect of the condition on the proportion of the robot interrupting the participant. The \textit{Restricted GUI} condition had the highest log-odds of destructive overlaps, corresponding to the probability of 17.8\%. In the \textit{Unrestricted GUI} condition, the log-odds reduced slightly, but this change was not significant \((p > 0.05)\). In the \textit{VR Interface} condition, the log-odds decreased significantly compared to both the \textit{Restricted GUI} and the \textit{Unresticted GUI} \((p < 0.001)\). 


\section{Discussion}
We conducted a mixed-design study where participants interacted with the Furhat robot as wizards controlling it or as users conversing with it. We investigated the differences in perceiving the interactions powered by three WoZ interfaces: Restricted GUI, Unrestricted GUI, and VR Interface, each offering a varying level of input and output capacity, from both the users' and the wizards' perspectives.

Regarding the users, we found a clear trend of \textit{VR Interface }being the most preferred condition and \textit{Restricted GUI} being the least preferred, as was shown by virtually every dimension of the Godspeed and Social Presence questionnaires and users' open-ended reflections. The \textit{Unrestricted GUI} was also clearly in the middle between the two conditions. Analyzing the speech parameters of the conversations, we found that in the \textit{Restricted GUI} condition, the robot was taking the longest pauses before producing an answer and was more likely to interrupt the participant. These parameters might have been the driving force of the subjective rating, as reflected in the open-ended responses. It also contained the least speech produced by the participant and relatively low speech from the robot, suggesting that the conversations in the \textit{Restricted GUI} condition contained a lot of silence, leading to disengagement and alienation. Notably, in the \textit{VR Interface} condition, the robot produced the least speech, while participants in this condition spoke the most. Given that the \textit{VR Interface} condition was most preferred by the participants, we suggest that timely and context-appropriate robot's responses encouraged the participants to be more engaged, while also enjoying the social qualities of the robot, despite limited verbal output. Turn-yielding and turn-processing gaps duration gradually increased from \textit{VR Interface} condition to \textit{Restricted GUI}. This not only suggests that the \textit{VR Interface} condition induced the most naturalistic conversational flow, but also might suggest that participants adapted to the robot's pace in all conditions, indicating prosodic synchrony \cite{sadoughi2017creating}.

As for the wizards' perspective, descriptive statistics suggest that the task load appears to be higher using the VR interface, both from NASA-TLX and SUS results. The wizards' open-ended reflections generally support this trend. This trend can be due to the selected scenario, as several wizards mentioned increased demand of remembering points of discussion. While none of the wizards mentioned experiencing motion sickness using the VR interface, the not-perfect latency and possible discomfort of wearing the VR headset could have also affected the wizards' preferences. Since the operators only had a 1-hour practice prior to the experiment, which is unusual for WoZ setups, most of the downsides identified in the telepresence condition can probably be faced by having a better-trained operator who is more experienced with the dialogue scenario and is more comfortable with socially interacting in VR. Despite that, several wizards reported feeling significantly more socially engaged in the \textit{VR Interface} condition, however for some it resulted in feeling more responsibility or concentration while performing their role. Notably, the wizards' preferences appear to vary, with some wizards having comfortable load levels and preferring the VR interface to the GUIs. The downsides that were pointed to the prototypical WoZ GUIs are harder to overcome and were already clearly identified in the related work section (hard to time turn-taking and limited flexibility in dialogue). Difficulty in timing and restriction in spontaneous reactions to the user was perceived as frustrating or socially awkward, with some wizards feeling rude to accidentally interrupt the user. The Restricted GUI was clearly the least preferred both by the wizards and the users as it exacerbates these issues whereas having non-restricted perception but restricted output appears to be the least demanding task for wizards.

\section{Limitations}
The small sample size, particularly of wizards, and their limited training time may have influenced the results.  Future research with a larger and more diverse sample, as well as more extensive training, is needed to further validate these findings. Also, other techniques and interfaces for achieving unrestricted output could be tested.




\section{Conclusion}
In this study, we compared the perception of interactions with a robot powered by three different WoZ interfaces by both users and wizards. 
Our VR-mediated interface demonstrated superior performance, was the most preferred, and provided a higher sense of social presence for users. It also fostered a greater sense of social connection for wizards, as reflected in their statements, such as “I really tried to perform” and acknowledgment of a “higher degree of responsibility” when operating the robot. Notably, turn-yielding gaps improved significantly using both VR interface and the Unrestricted GUI, highlighting the critical role of flexible input in achieving smoother turn-taking dynamics. 

The VR telepresence interface offers a promising path toward creating more natural and engaging social human-robot interactions. This is particularly important with the rapid advancements in generative AI, which are expected to enable robots to provide more responsive and adaptive feedback. As speech technology continues to evolve, testing interaction dynamics in ecologically valid ways is crucial. VR telepresence systems facilitate the collection of rich, contextualized data, paving the way for automating social behaviors in robots. While restricted input or output may still be valuable for controlled WoZ experiments, our findings highlight the trade-offs between different interface approaches.



\section*{Acknowledgments}
This work was supported by KTH Digital Futures (Sweden) and the Swedish Research Council project 2021-05803.

\bibliographystyle{ieeetr}
\bibliography{sample_base,refs}

\end{document}